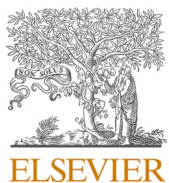
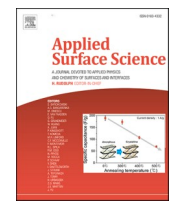

Full Length Article

# Into the origin of electrical conductivity for the metal–semiconductor junction at the atomic level

A. Janas [a], W. Piskorz [c], A. Kryshtal [b], G. Cempura [b], W. Belza [a], A. Kruk [b], B.R. Jany [a,*], F. Krok [a]

[a] *The Marian Smoluchowski Institute of Physics Jagiellonian University, Lojasiewicza 11, PL 30-348 Krakow, Poland*
[b] *Faculty of Metals Engineering and Industrial Computer Science and the International Centre of Electron Microscopy for Materials Science, AGH University of Science and Technology, PL 30-059 Krakow, Poland*
[c] *The Faculty of Chemistry, Jagiellonian University, ul. Gronostajowa 2, PL 30-387 Krakow, Poland*

ABSTRACT

The metal–semiconductor (M-S) junction based devices are commonly used in all sorts of electronic devices. Their electrical properties are defined by the metallic phase properties with a respect to the semiconductor used. Here we make an in-depth survey on the origin of the M-S junction at the atomic scale by studying the properties of the AuIn$_2$ nanoelectrodes formed on the InP(0 0 1) surface by the *in situ* electrical measurements in combination with a detailed investigation of atomically resolved structure supported by the first-principle calculations of its local electrical properties. We have found that a different crystallographic orientation of the same metallic phase with a respect to the semiconductor structure influences strongly the M-S junction rectifying properties by subtle change of the metal Fermi level and influencing the band edge moving at the interface. This ultimately changes conductivity regime between Ohmic and Schottky type. The effect of crystallographic orientation has to be taken into account in the engineering of the M-S junction-based electronic devices.

## 1. Introduction

Electronic devices based on metal–semiconductor (M-S) junction were one of the earliest electronic devices. Controlling of the electrical properties of the M-S junctions is critical because in all currently available electronic devices the electrodes are made of metal [1–4]. The metal–semiconductor junction is formed when a metal is brought into contact with a semiconductor material. In the very simple scenario, the M-S junction could possess rectifying or non-rectifying properties depending on the electronic properties of the materials it consists of, i.e. the work function of the metal and the electron affinity (*n*-type) or the ionization energy (*p*-type) of the semiconductor [5–7]. The non-rectifying M-S junction is called the Ohmic contact while the rectifying one – the Schottky diode or the Schottky contact. The behaviour of the rectifying properties of the M-S junction, discovered by Braun [8], was explained by Schottky [9] by introducing the so-called effective Schottky barrier (potential energy barrier) formed at the M-S interface [29–31]. The atomic arrangement at the M-S interface plays an essential role in defining its electrical properties. The simplest way to distinguish between the Ohmic contact or the Schottky diode is to analyze the Current-Voltage (*I-V*) characteristics of the device. The *I-V* relationship of the Ohmic contact is linear and follows the Ohm's law: $V = I \cdot R$, where *V* is the applied voltage, *I* – the flowing current, and *R* – the device resistance. Usually, also the specific contact resistance can be calculated as $r_c = RA$, where *A* is the active area. The *I-V* characteristic of the Schottky contact is non-linear and follows the thermionic emission equation [10] $I = AA^*T^2\exp\left(\frac{-\Phi_B}{k_BT}\right)\left[\exp\left(\frac{e_0(V-R_SI)}{\eta k_BT}\right) - 1\right]$, where *V* – applied voltage, *I* – flowing current, $R_S$ – series resistance, *T* – temperature, $k_B$ – Boltzmann constant, *A* – active area, $A^*$ – effective Richardson constant, $\eta$ – ideality factor, $\Phi_B$ – Schottky barrier height. This is true for the common semiconductors (Si, GaAs, GaN, InP, etc.) with high mobility where the barriers are not so thick, so the drift diffusion could be neglected. The M-S junction is usually characterized by the effective parameters obtained by fitting the *I-V* dependence to either the Ohm's Law or to the thermionic emission equation. These approaches characterize the electrical properties of the junction quite effectively without going into detail at the atomic scale.

Here, we present a comprehensive study, at the atomic scale, of the M-S junction formed between the Au-rich nanoelectrodes grown on the InP(0 0 1) single crystals, by the combination of the Conductive AFM (C-AFM) technique together with the atomically resolved High Angle

Annular Dark Field Scanning Transmission Electron Microscopy (HAADF STEM) measurements corroborated by the Density Functional Theory (DFT) calculations. The structurally characterised junction together with the Local Density of States (LDOS) and the C-AFM electrical measurements allowed us not only to effectively describe and understand the formed M-S junction by deriving its parameters, but also to depict it quantitatively at its origin, i.e at the atomic interface. Additionally, we see that the crystallographic orientation of the metal with respect to the semiconductor plays an essential role and defines the M-S junction rectifying or non-rectifying character by changing the nanoelectrode Fermi level and band edge moving at the interface.

## 2. Results and discussion

The AIII-BV semiconductors, in particular InP used in optoelectronic applications [12] or as a field-effect transistor (FET) based biosensor [13], are considered promising materials to overcome the limitations of the silicon-based technology. In all these device applications usually the Au-rich nanoelectrodes are used to provide the electric contact between the ambient and the fabricated device. It is hence important to study and to understand the electrical performance of the nanoelectrodes since they can influence also the performance of the final device. In the present study, the Au-rich nanoelectrodes made of $AuIn_2$ alloy were formed in the process of thermally induced self-assembly [11] of Au deposited by MBE on InP(0 0 1) n-doped single crystals. After samples preparation, the nanoelectrodes were electrically characterized *in situ* (in UHV) with C-AFM measurements. The *I-V* data were collected in a hyperspectral mode and the resultant map which shows the regions with the same *I-V* characteristics, was obtained together with average *I-V* characteristics in that region. The HAADF STEM measurements were performed on Focused Ion Beam (FIB) prepared sample's cross-sections. The Density Functional Theory (DFT) calculations of electronic properties, i.e. the Local Density of States (LDOS), were performed with the use of the VASP [17] code (for details please look into Methods Section). In Fig. 1a), the SEM morphology of $AuIn_2$ nanoelectrodes, formed on InP(0 0 1) surface, of 20–30 nm in diameter, is shown. A high-resolution AFM imaging, as depicted in Fig. 1(b–d), has shown that the $AuIn_2$ nanostructures are of two types of morphology. Some of the nanostructures exhibit fewer side facets with "Flat Top" (Fig. 1c) and the other facet-ones with "Sharp Top" (Fig. 1d). The C-AFM *I-V* hyperspectral data with simultaneously collected sample topography are presented in Fig. 1(e and f) (see also Supporting Information visualisation of the *I-V* data cube as a movie). The average current map in Fig. 1(e) clearly shows that there are nanoelectrodes which exhibit higher conductivity (higher average current) and lower conductivity (lower average current). The grouping (clustering) of the collected *I-V* hyperspectral data shows in detail the three different regions in terms of electrical properties as presented in the map (Fig. 1g) and corresponding average *I-V* of these regions presented in Fig. 1(h). The first region (blue colouring) of a lower conductivity corresponds to the nanoelectrodes which are of "Flat Top". This sample region exhibits the non-linear *I-V* characteristics. The second region (marked with red) of a higher conductivity corresponds to the area of nanoelectrodes with "Sharp Top" and exhibits the linear *I-V* characteristics. The third region (green) corresponds to the InP surface which shows typical for a semiconductor material, non-linear *I-V* behaviour. From our C-AFM measurements it has been found that ~70% of all AuIn$_2$ nanoelectrodes are of lower conductivity.

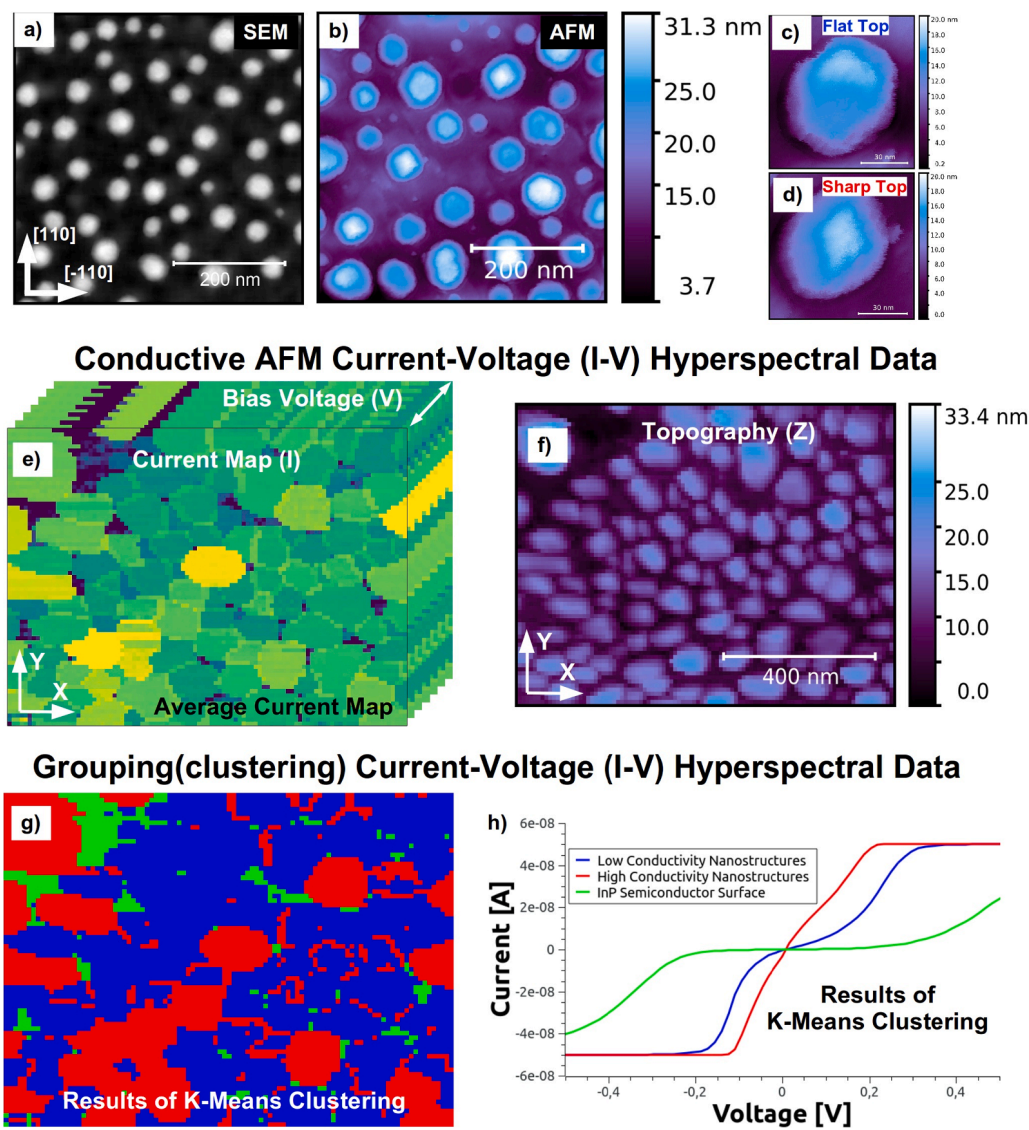

**Fig. 1.** Characterization of morphology and electrical properties of $AuIn_2$ nanoelectrodes grown on n-doped InP(0 0 1) surface. (a) SEM and (b) AFM topographies of the nanoelectrodes. Two different types of nanoelectrodes are visible: (c) "Flat Top" and (d) "Sharp Top". Results of C-AFM Current-Voltage (I-V) hyperspectral measurements performed *in situ* after nanoelectrodes synthesis: average current map (e) and corresponding topography (f). Low conductivity "Flat Top" and High conductivity "Sharp Top" nanoelectrodes are visible. Results of grouping (clustering) by k-means of current–voltage (I-V) hyperspectral data: map showing different I-V regions g) together with corresponding average I-V characteristics in these regions h). Three different regions are visible: low conductivity nanoelectrodes region (blue) which exhibit nonlinear I-V behavior, high conductivity nanoelectrodes region (red) with linear I-V characteristic, InP surface region (green). It is seen that ~70% of all nanoelectrodes are of lower conductivity. (For interpretation of the references to colour in this figure legend, the reader is referred to the web version of this article.)

To investigate the details of the AuIn$_2$ nanostructure/substrate interface, the atomic scale HAADF STEM measurements were performed. Fig. 2(a) and (b) show the HAADF STEM image of "Flat Top", low conductivity nanoelectrodes together with details of the *I-V* measurement results. The *I-V* data were fitted by thermionic Schottky equation, and the effective parameters of this M-S junction were extracted (see Fig. 2b). The Schottky barrier height of $\Phi_B = (0.2555 \pm 0.0080(stat) \pm 0.014(syst))eV$ and the ideality factor $\eta = (2.22 \pm 0.31(stat) \pm 0.12(syst))$ were obtained. It can be noted that the fitted Schottky barrier height is very similar to the pure indium to InP contacts which is in the order of 0.32 eV [19]. It can be seen from the HAADF STEM image that the (1–1 1) crystallographic plane of the formed AuIn$_2$ nanoelectrode is exposed towards the substrate. Resulting in the following crystallographic orientation (1–1 1)AuIn$_2$//(0 0 1)InP and [1 1 0]AuIn$_2$//[1 1 0]InP. Detail information on the crystal structures of the metal (AuIn$_2$) and InP is presented in Fig. S3 in Supporting Information. In Fig. 2c the atomically resolved AuIn$_2$/InP interface, with indium and phosphorus atomic columns clearly resolved, is depicted and overlaid with the DFT calculated atomic structural model. The model uncovers in detail the structure of the AuIn$_2$/InP interface. In the interface, the last layer of the substrate consists of In-P dimers with structurally disturbed positions of In atoms along the column (see Fig. 2c). On the other side, at the bottom of the nanoelectrode there is only one layer of Au-In dimers (see Fig. 2c) which are organized differently that the rest of AuIn$_2$ nanoelectrode arranged as linear trimers with Au atoms in their centre. The interface structure is not a perfect crystalline structure but rather a disturbed one, what is also reflected in the HAADF image. However, we do not observe any misfit dislocations or strain fields on the both, i.e. the nanoelectrode and substrate side. Furthermore, the local electronic properties of the M-S junction were characterised with the DFT calculations providing the computed LDOS along the direction normal to the AuIn$_2$/InP interface (see Fig. 2(d) and (e)). The LDOS colour map (Fig. 2d) composed of the slices reflecting the average electronic structures of subsequent atomic monolayers, shows the changes of electronic structure across the studied interface. To see the details of the electronic structure at the vicinity of the M-S junction interface we looked at the LDOS slices as from the DFT calculations (Fig. 2e). The InP LDOS is presented for three different regions, i.e., the bulk InP, the vicinity of the interface, and at the interface. For the comparison, also the LDOS for the AuIn$_2$ metal alloy is presented. The LDOS is calculated relative to vacuum at 0 eV. The valence band and the conduction band states of the InP can be clearly seen. The DFT calculated Fermi energy of AuIn$_2$ metal alloy in (1 1 1) configuration of $E_F = 4.247$ eV is presented as a red dashed line in Fig. 2(d)–(e).

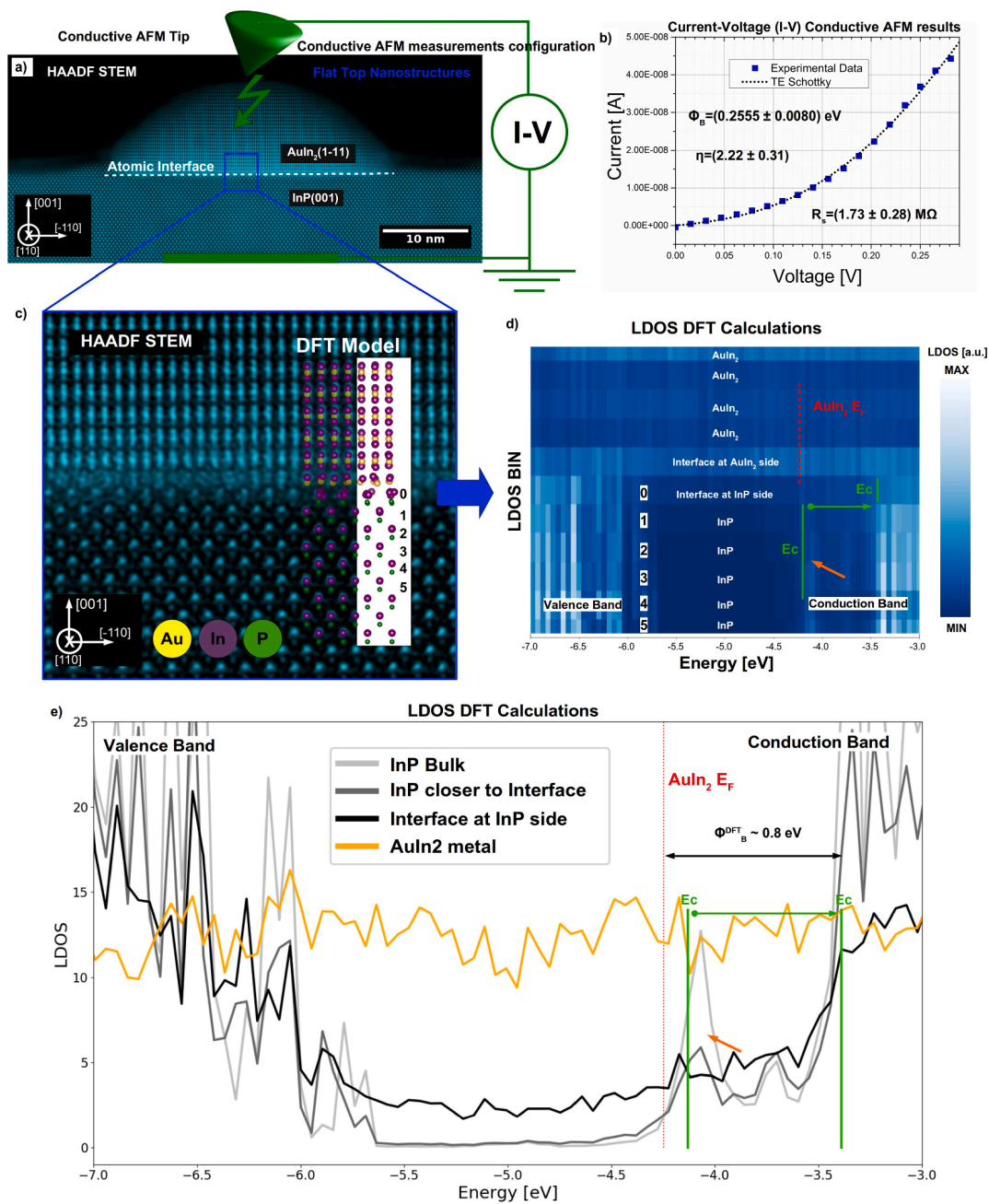

**Fig. 2.** Low conductivity, "Flat Top" AuIn$_2$ nanoelectrode grown on n-doped InP(0 0 1) surface. Atomically resolved HAADF STEM of the AuIn$_2$/InP system (a). Corresponding C-AFM Current-Voltage results with fitted Schottky barrier of 0.2555 eV according to the TE Schottky equation. HAADF STEM enlarged view into the interface (c) together with atomistic structural model from DFT calculations. DFT Local Density of States (LDOS) color map along the direction normal to the AuIn$_2$/InP interface (d). The LDOS of InP bulk region and the regions close to the interface together with LDOS of AuIn$_2$ electrodes relative to the vacuum energy at 0 eV e). It is seen that the Conduction Band Edge (Ec) moves as one approaches the interface region. The LDOS state ~4 eV melts completely at the interface (see green arrow). Comparing Ec at interface with the Fermi energy of AuIn$_2$, the Schottky barrier of ~0.8 eV was approximated from DFT.

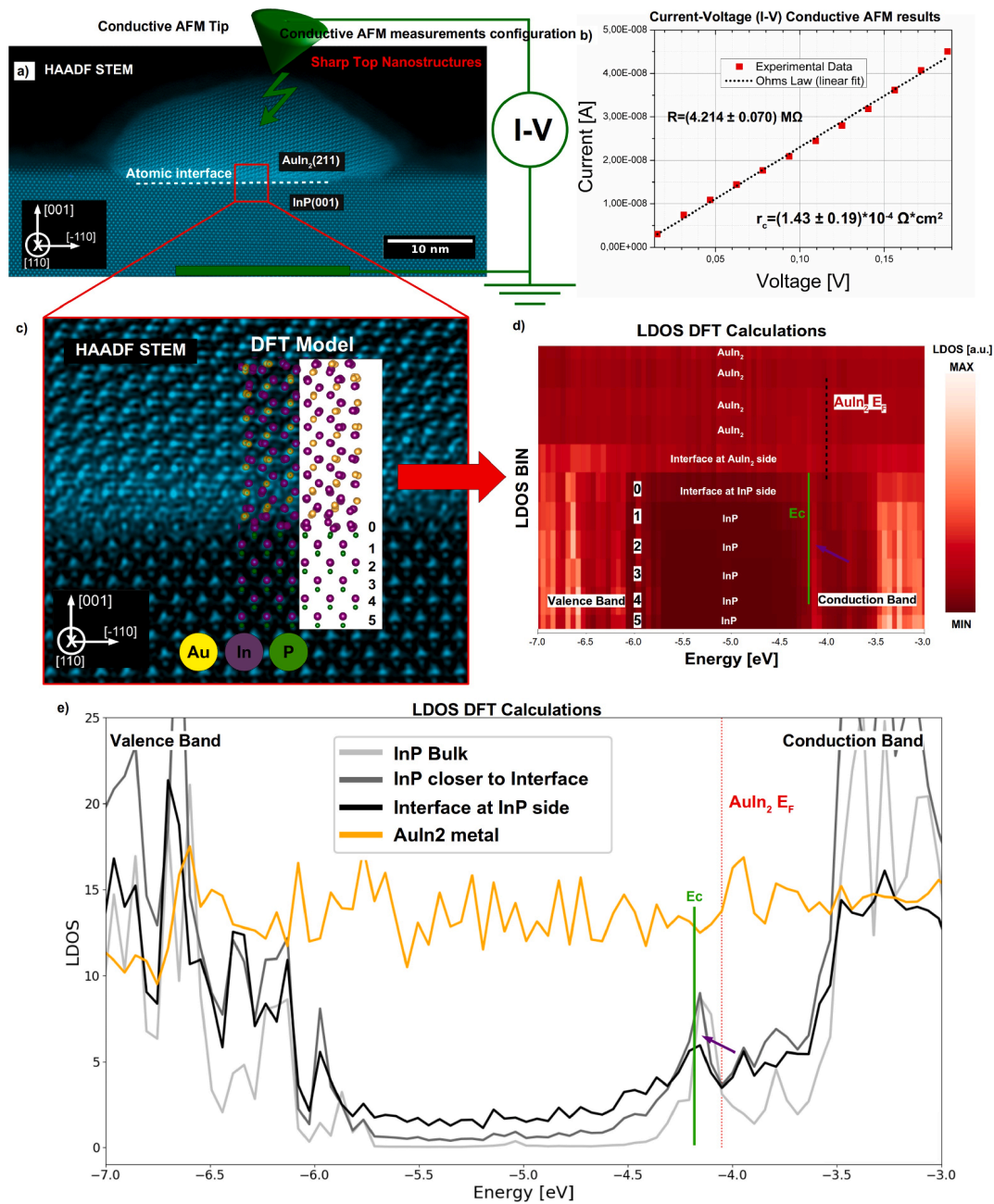

**Fig. 3.** High conductivity AuIn$_2$ nanoelectrodes "Sharp Top" on n-doped InP(0 0 1) surface. Atomically resolved HAADF STEM of high conductivity nanoelectrode (a). Corresponding C-AFM Current-Voltage results together with Ohm's law linear fit. HAADF STEM enlarged view into the interface (c) together with atomistic structural model from the DFT. Calculated by DFT Local Density of States (LDOS) color map along the direction normal to the AuIn2/InP interface (d). DFT calculated LDOS of InP bulk region and the regions close to the interface together with LDOS of AuIn$_2$ electrodes relative to the vacuum energy at 0 eV (e). It is seen that the Conduction Band Edge (Ec) does not move as one approaches the interface region. The LDOS state at ~4 eV does not melt at the interface (see green arrow). No Schottky barrier is observed. (For interpretation of the references to colour in this figure legend, the reader is referred to the web version of this article.)

It can be noticed that the local band structure of InP is changing as one approaches the interface with the AuIn$_2$ nanoelectrode, i.e., the Conduction Band Edge ($E_c$) of InP is moving towards the vacuum level (energy of 0 eV). This is related to the disturbed atomic structure of the Interface at InP side, as can be inferred from the DFT model (Fig. 2c). This disturbed structure induces the localized states at the band gap region, so the band gap and Conduction Band Minimum (CBM) cannot be unambiguously defined, in contrast to the Conduction Band Edge ($E_c$), similarly as for the amorphous material case [21]. The effective $E_c$ movement is realized by a total melting of the InP LDOS state at ~4 eV (see DFT LDOS in Fig. 2(d) and (e)). This LDOS melting effect extends over around three InP atomic layers. Since now there are almost no InP LDOS states at metal Fermi energy level, the flowing electron current from the nanoelectrode toward InP will effectively feel the barrier at the interface region. From the difference between the $E_c$ of InP at the interface and the Fermi Energy of AuIn$_2$ the value of the Schottky barrier height can be approximated as ~0.8 eV. This simple approximation, stemming from the DFT calculations, is different from the experimental value due to the calculation of the exchange–correlation energy which is a well known effect [20]. This, however, agrees qualitatively with the experimental results which show the appearance of the Schottky behaviour.

We now analyse the high conductivity AuIn$_2$ nanoelectrodes. Fig. 3 (a) and (b) show the HAADF STEM image of the high conductivity "Sharp Top" nanoelectrodes together with the details of the $I$-$V$ measurements. The linear $I$-$V$ data were fitted to the Ohm's law and the specific contact resistance was extracted: $r_c = (1.43 \pm 0.19(stat) \pm 0.20(syst))*10^{-4}$ $\Omega cm^2$. This value is consistent with a bulk gold Ohmic contacts to InP [19]. It can be seen from the HAADF STEM image that these AuIn$_2$ nanoelectrodes expose a (2 1 1) crystallographic plane towards the InP(0 0 1) substrate surface, which is different compared to the case of the low conductivity nanoelectrodes. This results in the crystallographic orientation of (2 1 1)AuIn$_2$//(0 0 1) InP and [1–1 1]AuIn$_2$//[1 1 0]InP. The atomically resolved HAADF STEM image of the AuIn$_2$/InP interface is shown in Fig. 3(c) together with the DFT calculated atomic structural model of this M-S junction. The interface region at the InP side consists of the In-P dimers with disturbed atomic structure in columns. While interface at the AuIn$_2$ side consists of the In-Au-In trimers, also with the disturbed structure in atomic columns. The HAADF contrast blur indicates a reduced atomic order of the interface. However, we do not observed any misfit dislocations or strain fields on the both, i.e. the nanoelectrode and substrate side. The calculated LDOS along the direction normal to the AuIn$_2$/InP interface in presented Fig. 3(d)–(e). The LDOS colour map Fig. 3(d) shows similarly how the local electronic structure changes when the atomic structure changes from InP to AuIn$_2$. The detailed LDOS slices are presented in Fig. 3(e). The DFT calculated Fermi Energy of AuIn$_2$ metal alloy in the (2 1 1) configuration of $E_F = 4.051$ eV is presented as a red dashed line in Fig. 3(d)–(e). It can be noticed that the local band structure is not changing as one approaches the interface, i.e. the Conduction Band Edge (Ec) of InP is not moving. This time the LDOS state at ~4 eV does not melt as shown in Fig. 3(d) and (e). We have, therefore, no Schottky barrier and the contact is fully Ohmic. This agrees with the experimental results which show the Ohmic behaviour.

It is now important to note that the difference between these two nanoelectrodes is the crystallographic orientation of the metal AuIn$_2$ phase with respect to the InP substrate resulting in one case, in formation of a Schottky-type junction and, in another one, the Ohmic contact.

The nanoelectrodes were formed during self-assembling process, where during ad-atoms diffusion and aggregation the AuIn$_2$ nanoelectrode grows on InP surface by the formation of the most energetic favourable planes, having low surface energies i.e. (1 1 1) and (2 1 1) planes, which are very common for metals with cubic crystal structure [28]. This finally results in a two types of nanoelectrodes. The effect of the formation of Schottky nanodiode or Ohmic nanocontact is related to the crystallographic orientation implying changes in the Fermi level of the AuIn$_2$ metal alloy phase at the interface. The particular atomic structure at the interface provide the differences in LDOS electronic states and, consequently, is responsible for the appearance of the band edge movement, or the lack thereof, at the interface. We also see that in the case of the Ohmic-type junction the atomic structure, as seen from a top surface view, of InP and AuIn$_2$ is better matched together than in the case of Schottky-type junction, see Fig. S4 in Supporting Information. This is also directly seen in the calculated lattice misfit, see Table S1 in Supporting Information. It is seen that on average the lattice misfit is lower for high conductivity nanoelectrodes.

What we also see is that in case of the studied here AuIn$_2$/InP(n-type) the change in M-S junction behaviour between the Ohmic and Schottky regimes is significant since the work function ($E_f$) of metal nanoelectrode (AuIn$_2$) is close to the Conduction Band Minimum (CBM), for p-type semiconductor this will be when $E_f$ will be close to the Valence Band Maximum (VBM). The work function ($E_f$) changes induced by crystallography are significant to allow here for the switching between the Ohmic and Schottky conduction regimes. As we think, this effect could be also applicable to other semiconductor systems. The corresponding energy relations between the metal $E_f$ and substrate semiconductors' CBM and VBM are presented in Fig. 4. By comparing CMB (n-type) or VBM (p-type) with a work functions for the selected metals [24–26], we proposed the M-S systems where the effect of crystallographic orientation of metal with a respect to the semiconductor will cause a changes in conductivity behaviour between Ohmic and Schottky regime. The conductivity behaviour changes of the M-S junction related to the crystallographic orientation of the metal are a consequence of the metal work function changes at the interface. This effect could be used to tune the parameters of M-S junction to desired behaviour within a single metal or alloy phase by controlling the growth of a metal with a desired orientation on semiconductors substrates using various advanced heteroepitaxy methods.

## 3. Conclusions

Based on the studied AuIn$_2$ nanoelectrodes which form two types of M-S junctions, i.e. the Schottky or the Ohmic contacts, on InP(0 0 1) substrate surface, we have shown directly at atomic scale that the different crystallographic orientation of the same metal with respect to the semiconductor, thus different structure of the interface, determines the electrical properties of the M-S junction. Hence, the rectifying junction in one case, and the non-rectifying one in the other case, is formed by changing the Fermi level of the AuIn$_2$ metal alloy phase and influencing local changes in electronic structure, i.e. band edge moving at the disturbed interface. We directly see that the origin of this different conductivity behaviour has its roots in the mutual crystallographic orientation of the metal nanoelectrode and the semiconductor. The effect of the crystallographic orientation of the metal electrode with respect to the substrate, implying the Fermi level changes at the interface, can be used to control the electrical properties of the M-S junction based devices by utilizing only single metal/alloy phase to change between Ohmic-Schottky conductivity regimes. Our findings could be also applicable to other metal-semiconductors systems. This gives a possibility for the engineering of the new desired future electronic devices using metal electrodes with defined optimally suited crystallographic orientation in particular in the area of AIII-BV based devices.

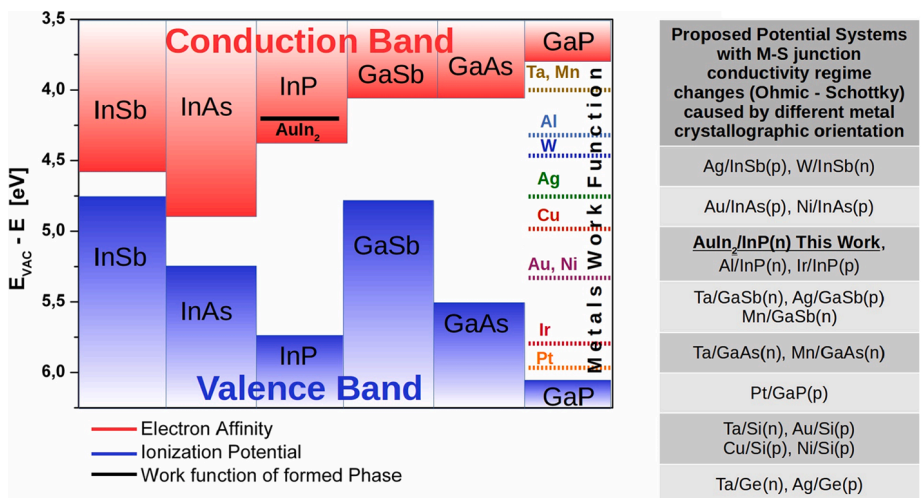

**Fig. 4.** Energy relative to the vacuum as a function of electron affinity (CBM) and ionization potential (VBM) for AIII-BV semiconductors. Work functions of selected metals are presented as a horizontal lines. Work function for the studied here $AuIn_2$ on InP is also presented as a black line. Based on our work on $AuIn_2$/InP system, by comparing CBM for n type semiconductors and VMB for p type semiconductors with a metals work function ($E_f$), the table which presents potential systems, where M-S junction conductivity regime changes (Ohmic–Schottky) caused by different metal crystallographic orientation is derived.

## 4. Methods

### 4.1. Sample preparation

Indium phosphide InP(0 0 1) n-doped crystal was mounted on molybdenum plate and introduced into the UHV Molecular Beam Epitaxy (MBE) system, with a base pressure of $10^{-10}$ mbar. The sample was initially out-gassed for 1 h at 150 C and exposed to low energy, 700 eV Ar + ion bombardment at a 60 deg incident angle at room temperature (RT). The sample surface was cleaned in cycles of ion-beam irradiation at T = 450 C (InP) until the (4 × 2) InP(0 0 1) reflection high energy diffraction (RHEED) pattern was observed (see Supporting Information Fig. S1). The applied cleaning procedure results in atomically flat surfaces. Next, 2 ML (mono-layers) of Au was deposited on the sample surface at temperature of 330 C and a rate of 0.1 ML/min as checked with a quartz micro-balance. After the deposition, the sample was cooled down to RT at a rate of 10 C/min. The sample temperature during all experiments was measured with a pyrometer (LumaSense, model IGA 140) with emissivity ε = 0.5. The applied sample preparation results in the formation of the metallic nanoelectrodes on the sample surface made of $AuIn_2$ alloy [11].

### 4.2. Sample characterization

The electrical characterization of the nanoelectrodes in the form of current voltage (I-V) measurements was carried out *in situ* (in UHV) just after sample preparation by Conductive Atomic Force Microscopy (C-AFM) using Omicron RT AFM/STM microscope. The I-V data were collected in a hyperspectral mode in the form of the three dimensional Spectrum Image stack, where for each sample grid point (x, y) a full I-V curve was collected as z-axis. The C-AFM data were grouped together (clustered) using K-Means method as implemented in Scikit-Learn [14]. The resultant map, which shows the regions with the same *I-V* characteristics, was obtained together with average *I-V* characteristics in that region. To extract the Schottky barrier height and ideality factor, the collected I-V data by C-AFM were fitted by the thermionic emission equation [3,10] $I = AA^*T^2\exp\left(\frac{-\Phi_B}{k_BT}\right)\left[\exp\left(\frac{e_0(V-R_SI)}{\eta k_BT}\right) - 1\right]$, where $V$ – applied voltage, $I$ – flowing current, $R_S$ – series resistance, $T$ – temperature, $k_B$ – Boltzmann constant, $A$ – active area, $A^*$ – effective Richardson constant, η – ideality factor, $\Phi_B$ – Schottky barrier height. Next, the sample was transferred under ambient conditions to perform high-resolution Atomic Force Microscopy (AFM) imaging by AMF NanoScope microscope working in Peak Force mode from Bruker. The SEM imaging in secondary electrons (SE) mode was performed by Dual Beam SEM/FIB FEI Quanta 3D FEG microscope from FEI. Later, the atomically resolved HAADF STEM measurements, where the contrast is proportional to the atomic number Z and to the sample thickness, were performed using a FEI (S)TEM Titan3 G2 60-300 microscope operated at 300 kV. The HAADF-STEM images were acquired with a convergence angle of 20 mrad and a probe current of 80 pA. To get rid of the scanning artefacts the HAADF STEM data were collected as an image stack of ten 4 k HAADF STEM images, which next were registered by Non-rigid registration using free software ImageJ/FIJI [27] and median stacked. Before median stacking the registered images were scaled by a factor of four. The obtained HAADF-STEM images were deconvoluted to remove the overall blur caused by different effects (source size, aberration, other instabilities, etc.) and to increase the image resolution. We used the assumption of ideal microscope and ideal crystal structure in the bulk i. e. the ideal crystal structure in the ideal microscope will be visible as points. By this assumption the transfer function is derived, by the fit in the Fourier space, as an asymmetric Gaussian function, which includes all the blurring effects. The derived transfer function is later used for the deconvolution of the whole image. As we checked in details, such a deconvolution approach does not introduce any artefacts, the image resolution is increased making the interpretation of the structure easier. All the deconvolution approach steps were done by the free software Gwyddion [16]. For details please see Supporting Information Fig. S2. The STEM measurements were performance on the Focused Ion Beam (FIB) prepared thin foils from nanoelectrodes sample which was covered by thermally evaporated carbon layer in the UHV Chamber to prevent surface contamination and damage [15].

The quantum-chemical calculations of electronic properties i.e. the Local Density of States (LDOS), were performed by the Density Functional Theory (DFT) calculations with the use of the VASP [17] code. To reliably extract the information of the $AuIn_2$/InP atomic interface (which is not trivial), the atomic model was derived purely from DFT calculations, which used as an input only orientations of both structures, i.e. the nanoelectrode and the substrate, which are indisputably visible. Only then the optimized by DFT model is validated on the HAADF data. This kind of approach, which we used does not depend on the HAADF STEM interface data directly, which contains non trivial disturbed atomic structure. The DFT optimized model of the atomic interface delivered the same atomic structure as in the HAADF STEM data, which finally validated our approach. The derived full atomic model together with disturbed atomic structure was used for the electronic properties calculations. The Γ point sampling of the irreducible Brillouin zone was used for the $AuIn_2$/InP interface supercell LDOS calculations together with the Methfessel-Paxton smearing of 0.01 eV and the PBE [22,23] functional. The plane wave energy cut-off was chosen as 400 eV. In the DFT LDOS calculations, the electron affinity of pristine InP was calibrated to resemble the experimental value of 4.38 eV [18].

**Author contributions**

A.J. and B.R.J. contributed to the characterization of the samples by conductive AFM and RHEED/SEM together with AFM/RHEED/SEM/TEM data analysis and interpretation. A.J. prepared the samples in UHV. A.J., B.R.J., W.B. contributed to the C-AFM measurements in UHV. A.K., G.C., A.K. contributed to HAADF STEM measurements. B.R.J. contributed to the FIB sample preparation, to the analysis and interpretation of *I-V* data and to HAADF STEM image deconvolution. W.P. contributed to the DFT calculations. B.R.J. prepared the manuscript in consultation with F.K. and with all other authors. B.R.J. together with F.K. supervised and guided the measurements and analysis. B.R.J. initiated and organized this project.

**CRediT authorship contribution statement**

**A. Janas:** Investigation, Formal analysis. **W. Piskorz:** Software, Methodology, Writing – review & editing. **A. Kryshtal:** Investigation, Data curation. **G. Cempura:** Investigation, Data curation. **W. Belza:** Investigation. **A. Kruk:** Supervision. **B.R. Jany:** Conceptualization, Project administration, Supervision, Methodology, Investigation, Formal analysis, Data curation, Writing – original draft. **F. Krok:** Supervision, Writing – review & editing.

**Declaration of Competing Interest**

The authors declare that they have no known competing financial interests or personal relationships that could have appeared to influence the work reported in this paper.

**Acknowledgement**

B.R.J. acknowledges the help of Dr Konrad Szajna together with MSc Karol Cieslik during *in situ* in UHV C-AFM measurements of the nanoelectrodes. BRJ thanks Prof. Jozef Spalek and Prof. Rafal Abdank-Kozubski for the fruitful discussions about electrical properties of interfaces. F.K. acknowledges the support by the Polish National Science Center under UMO-2018/29/B/ST5/01406.

**Appendix A. Supplementary material**

Supplementary data to this article can be found online at https://doi.org/10.1016/j.apsusc.2021.150958.

# Supporting Information

## Into the Origin of Electrical Conductivity for the Metal-Semiconductor Junction at the Atomic Level

A.Janas[a], W. Piskorz[c], A. Kryshtal[b], G. Cempura[b], W. Belza[a], A. Kruk[b], B.R. Jany[a*], F. Krok[a]

[a]The Marian Smoluchowski Institute of Physics Jagiellonian University, Lojasiewicza 11, 30-348 Krakow, Poland
[b]Faculty of Metals Engineering and Industrial Computer Science and the International Centre of Electron Microscopy for Materials Science, AGH University of Science and Technology, PL 30-059 Krakow, Poland
[c]The Faculty of Chemistry, Jagiellonian University, ul. Gronostajowa 2, PL 30-387 Krakow, Poland

*Corresponding author e-mail: benedykt.jany@uj.edu.pl

# Nanoelectrodes sythesis and *in situ* formation dynamics

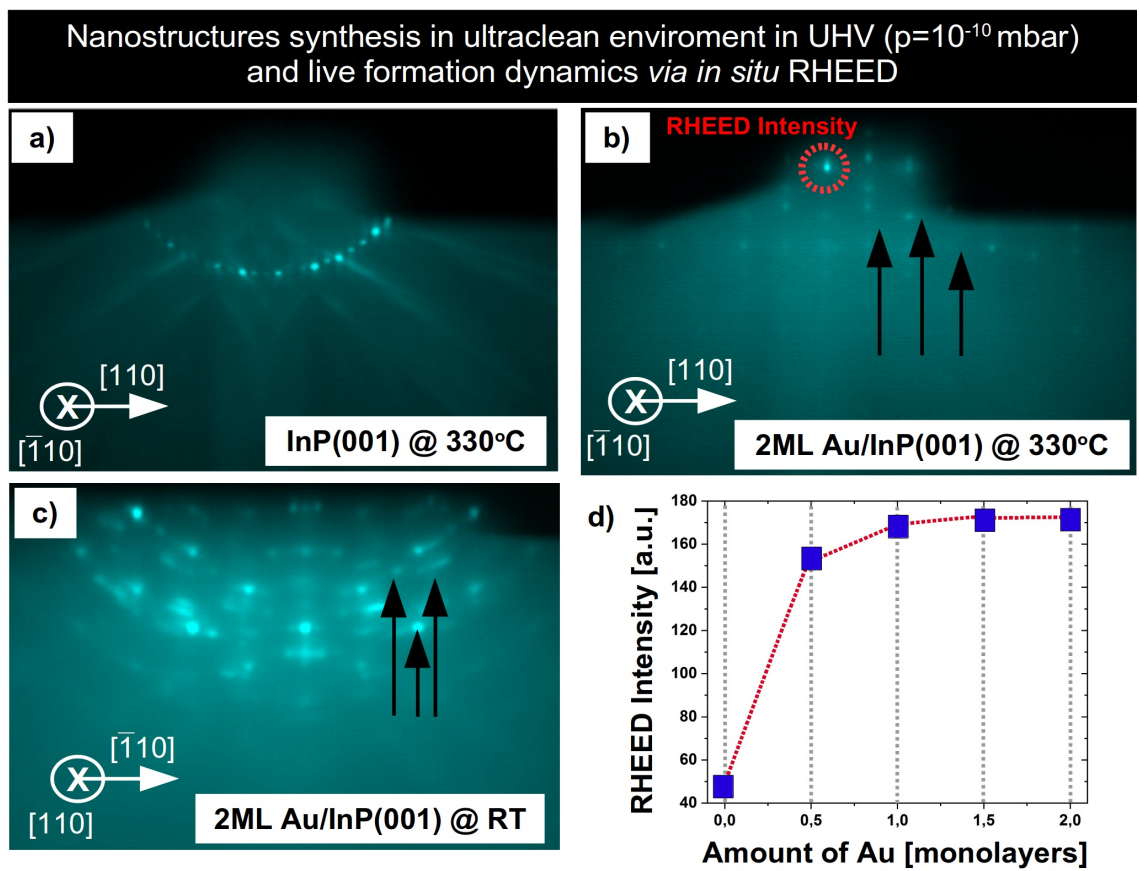

*Figure S1: Live dynamics of nanoelectrodes synthesis by in situ RHEED. RHEED pattern of the atomically clean and reconstructed InP(001) surface a). RHEED pattern on InP(001) surface after deposition of 2ML of Au at 330C b).Black arrows indicate diffraction spots from 3D structures i.e. nanoelectrodes. Red circle denotes diffraction spot used for the live dynamics investigation. RHEED patter of after sample cool down to room temperature c). RHEED intensity as a function of deposited amount of gold in monolayers d). Deposition of 0.5ML of Au results in appearance of RHEED pattern from 3D structures i.e. nanoelectrodes.*

Nanoelectrodes formation dynamics was studies during synthesis in UHV conditions by in situ RHEED diffraction analysis Figure S1. The RHEED shows that the InP(001) surface is before Au deposition is atomically clean and Figure S1a). The obtained results show that the nanoelectrodes start to from from 0.5ML of Au, where the first 3D RHEED pattern appears.

# HAADF STEM Image Deconvolution

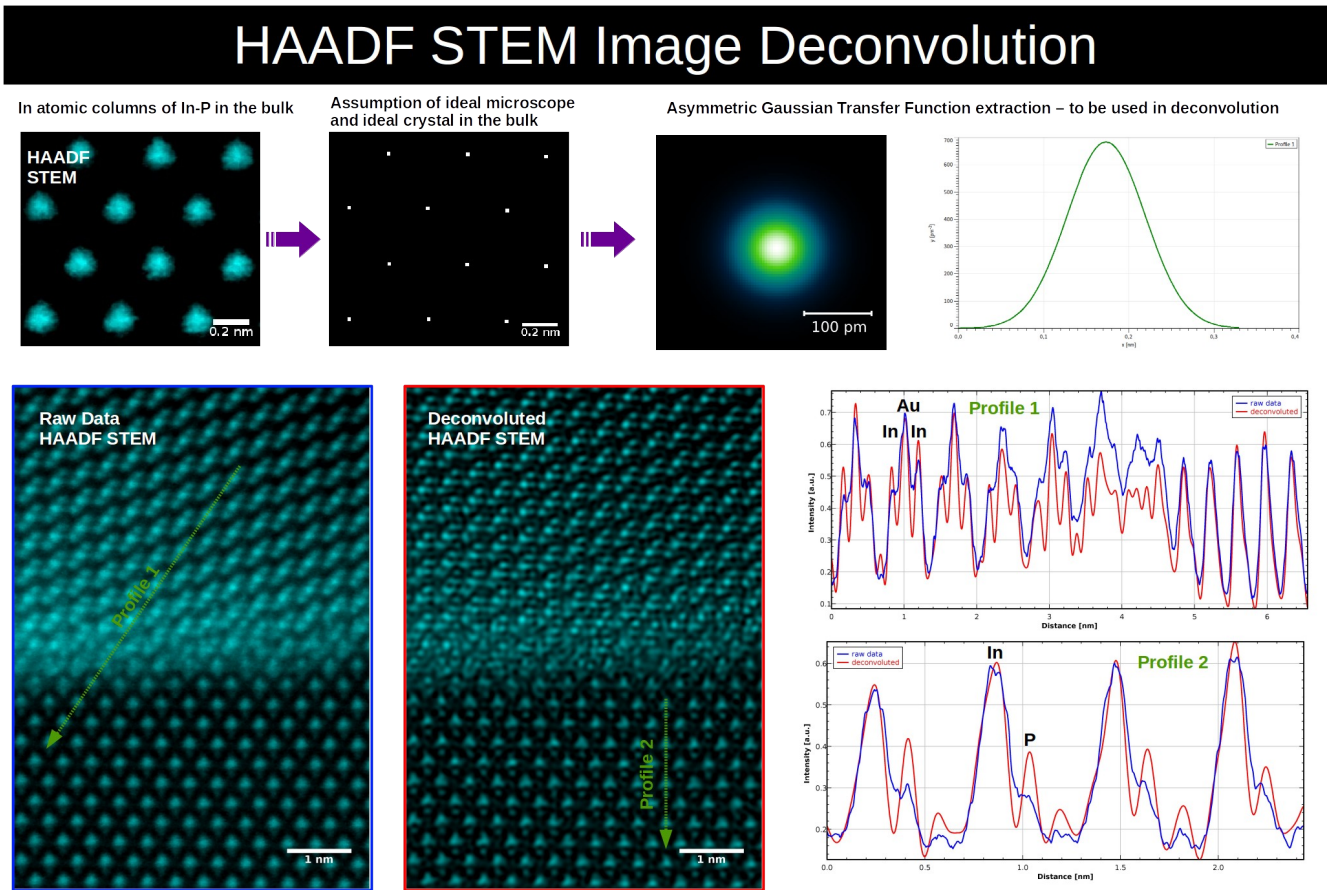

*Figure S2: HAADF STEM image deconvolution. Processing steps of deconvolution (top row): HAADF STEM of Indium atomic columns of In-P in the bulk, assumption of ideal microscope and ideal crystal lattice in the bulk in form of perfect points which represent perfect Indium atomic columns, extracted asymmetric Gaussian transfer function to be used in deconvolution. Comparison of deconvolution results (down row): raw HAADF STEM image, deconvoluted HAADF STEM image by using the extracted Gaussian transfer function (two line profiles indicated), comparison of raw HAADF STEM data with deconvoluted one by examining exactly the same line profiles. It is seen that on the deconvoluted data the atomic columns are much more clearly visible, the resolution is increased. As we see, such a deconvolution approach does not introduce any artefact's, the image resolution is increased making the interpretation of the structure easier. The whole processing steps as well as final deconvolution were done using free software Gwyddion (http://gwyddion.net).*

To get rid of the overall blur caused by different effects (source size, aberration, and other instabilities etc.) and to increase the image resolution, the obtained HAADF STEM images were deconvoluted using free software Gwyddion http://gwyddion.net (Nečas, D.; Klapetek, P. Gwyddion: An Open-Source Software for SPM Data Analysis. *Open Physics* **2011**, *10* (1), 181–188. https://doi.org/10.2478/s11534-011-0096-2.). First only Indium atomic columns were extracted from the In-P HAADF STEM in the bulk region Figure S2 (top row). Next the assumption of ideal microscope and ideal crystal lattice in the bulk was used i.e. the ideal crystal structure in the ideal microscope will be visible as points. The asymmetric Gaussian transfer function was extracted by the fit in Fourier space, which includes all the blurring effects (using "Statistics→Transfer Function Fit"). Later the extracted transfer function Figure S2 (top row) was used to deconvolve the collected HAADF STEM data (using "Multidata → Deconvolve"). High frequency noise was removed from deconvoluted image by performing standard 2D FFT filtering (using "Correct Data → 2D FFT Filtering"). The results of the deconvolution were compared with the original image by comparing exactly the same line profiles of the atomic columns Figure S2 (down row). It is seen that in the deconvoluted data the atomic columns are much more clearly visible, the resolution is increased. The Phosphorus atomic columns are much more clearly resolved. As we see in details, such a deconvolution approach does not introduce any artefact's, the image resolution is increased making the interpretation of the structure easier.

# Phases Crystallography

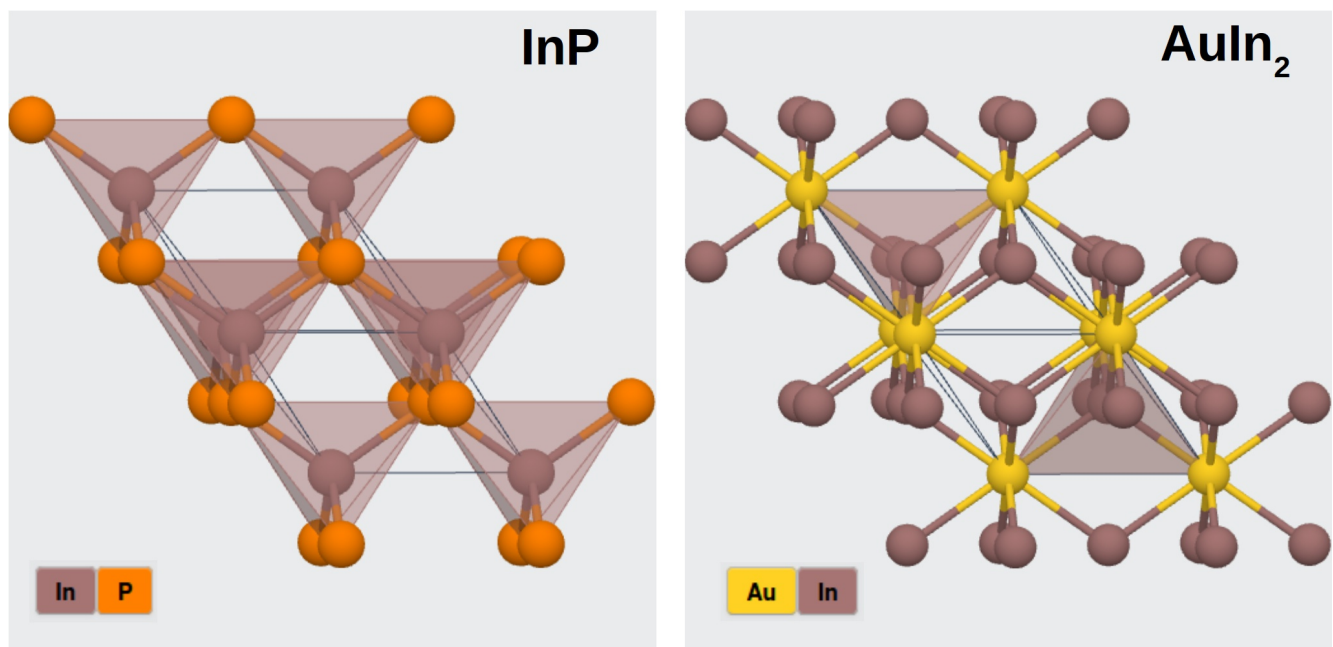

*Figure S3: Atomic structural models showing crystallography of InP and AuIn$_2$ alloy phases. Materials Project InP (doi:10.17188/1195475), AuIn$_2$ (doi:10.17188/1197382).*

In Figure S3 atomic structural models of InP and AuIn$_2$ phases are shown, Materials Project InP (doi:10.17188/1195475), AuIn$_2$ (doi:10.17188/1197382).

InP is Zincblende, Sphalerite structured and crystallizes in the cubic F-43m space group. The structure is three-dimensional. *In3+* is bonded to four equivalent *P3-* atoms to form corner-sharing *InP4* tetrahedra. All *In–P* bond lengths are 2.58 Å. *P3-* is bonded to four equivalent *In3+* atoms to form corner-sharing *PIn4* tetrahedra [1].

AuIn$_2$ is Fluorite structured and crystallizes in the cubic Fm-3m space group. The structure is three-dimensional. *Au* is bonded in a body-centered cubic geometry to eight equivalent *In* atoms. All *Au–In* bond lengths are 2.90 Å. *In* is bonded to four equivalent *Au* atoms to form a mixture of edge and corner-sharing *InAu4* tetrahedra [1].

[1] Ganose, A., & Jain, A. (2019). Robocrystallographer: Automated crystal structure text descriptions and analysis. MRS Communications, 9(3), 874-881. https://doi.org/10.1557/mrc.2019.94

# Nanoelectrodes atomic models

## Nanoelectrodes Surface – Top View

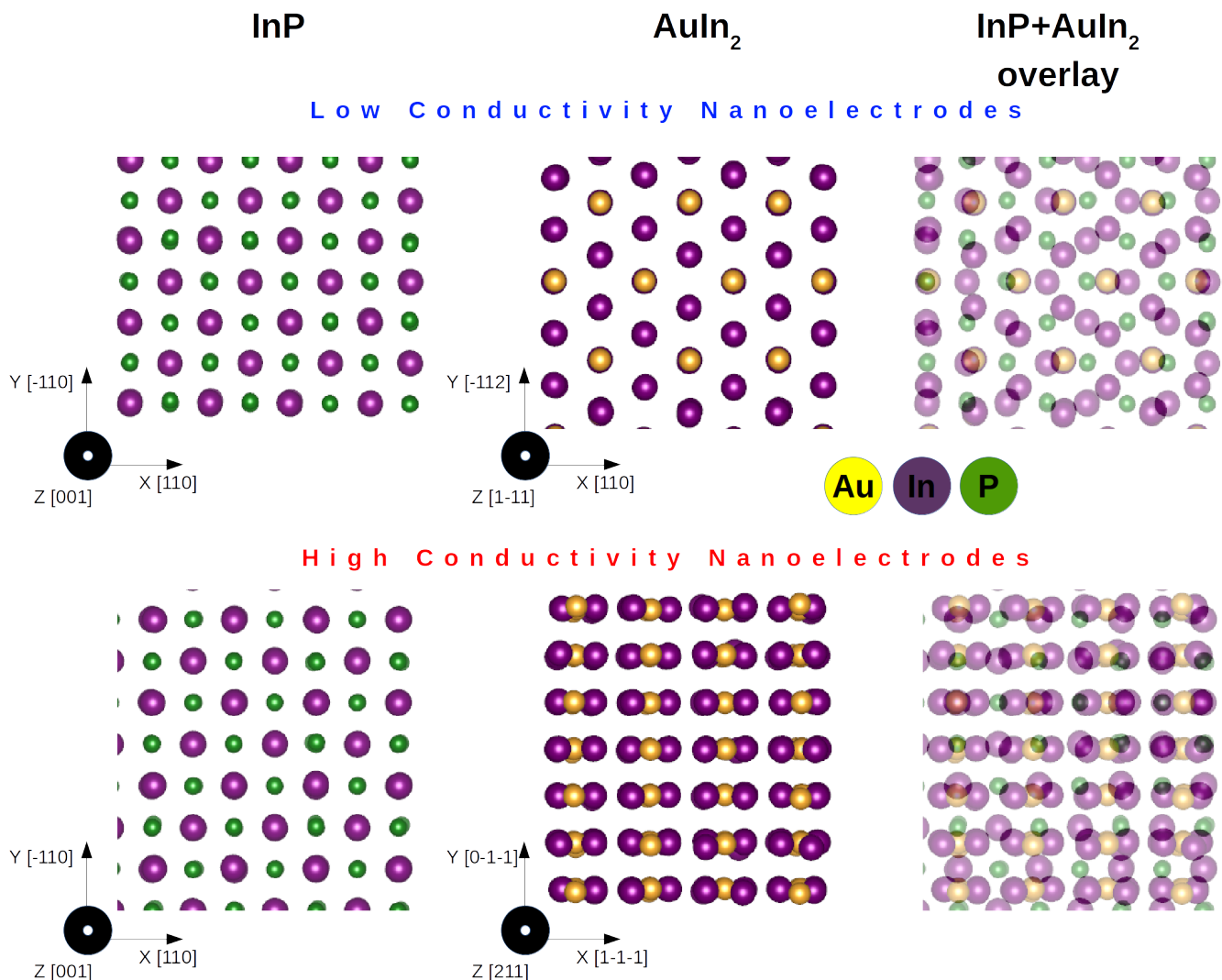

*Figure S4: Atomic structural models of a top surface view of AuIn$_2$ nanoelectrodes in respect to the InP(001) surface. Upper row shows atomic models of InP and AuIn$_2$ phase together with models overlay for low conductivity nanoelectrodes, lower row shows models for high conductivity nanoelectrodes respectively, crystallographic directions indicated. From the lattice overlay it is seen that the InP and AuIn$_2$ structures are better matched for high conductivity nanostructures.*

Figure S4 shows atomic structural models of a top surface view of AuIn$_2$ nanoelectrodes in respect to InP(001) surface. Models for InP and AuIn$_2$ phase are shown for two different crystallographic orientations corresponding to low conductivity nanoelectrodes (Figure S4 upper row) and to high conductivity nanoelectrodes (Figure S4 lower row). From the atomic models overlay (Figure S4 right column) it is seen that InP and $AuIn_2$ structures are better matched together for high conductivity nanoelectrodes. This is directly seen also in the calculated lattice misfit Table S1. It is seen that on average the lattice misfit is lower for high conductivity nanoelectreodes.

| Low Conductivity Nanoelectrodes | | | | |
|---|---|---|---|---|
| **InP *hkl*** | **d [A]** | **$AuIn_2$ *hkl*** | **d [A]** | **Misfit [%]** |
| "001" | 5.868 | "1-11" | 3.74 | 36.3% |
| "002" | 2.934 | "1-11" | 3.74 | 27.5% |
| "-110" | 4.15 | "-112" | 2.64 | 36.4% |
| "-220" | 2.075 | "-112" | 2.64 | 27.2% |
| "110" | 4.15 | "110" | 4.598 | 10.8% |
| High Conductivity Nanoelectrodes | | | | |
| **InP *hkl*** | **d [A]** | **$AuIn_2$ *hkl*** | **d [A]** | **Misfit [%]** |
| "001" | 5.868 | "1-11" | 2.64 | 55.0% |
| "002" | 2.934 | "211" | 2.64 | 10.0% |
| "-110" | 4.15 | "-112" | 4.598 | 10.8% |
| "110" | 4.15 | "110" | 3.74 | 9.9% |

*Table S1: Inter-planar spacing of InP and $AuIn_2$ phase together with corresponding miller indexes and calculated lattice misfit respectively, for low conductivity nanoelectrodes and for high conductivity nanoelectrodes. It is seen that on average the lattice misfit is lower for high conductivity nanoelectreodes.*